\documentstyle[12pt]{article}

\begin{document}

\newcommand{\beq}{\begin{equation}}
\newcommand{\eeq}{\end{equation}}
\newcommand{\E}{\mbox{\boldmath $E$}}
\newcommand{\B}{\mbox{\boldmath $B$}}
\newcommand{\C}{\mbox{\boldmath $C$}}
\newcommand{\Q}{\mbox{\boldmath $Q$}}
\newcommand{\bp}{\mbox{\boldmath $p$}}
\newcommand{\bd}{\mbox{\boldmath $d$}}
\newcommand{\cp}{\bar{\mbox{\boldmath $p$}}}

\def\F{\cal F}
\def\D{\cal D}
\def\a{\alpha}
\def\b{\beta}
\def\g{\gamma}
\def\d{\delta}
\def\p{\bar p}

\begin{titlepage}

\begin{center}

\hfill hep-th/9608118

\vskip 1.8cm
{\bf A Canonical Approach to Self-Duality of Dirichlet $3$-Brane}

\vskip .8cm

Shijong Ryang

\vskip .8cm
{\em Department of Physics \\ Kyoto Prefectural University of Medicine \\
Taishogun, Kyoto 603 Japan}

\end{center}

\vskip 2.0cm

\begin{center} {\bf Abstract} \end{center}
The self-duality of Dirichlet $3$-brane action under the $SL(2,R)$ duality
transformation of type IIB superstring theory is shown in the Hamiltonian
form of the path integral for the partition function by performing the
direct integration with respect to the boundary gauge field. Through the
integration in the phase space the canonical momentum conjugate to the
boundary gauge field can be effectively replaced by the dual gauge field.

\vskip 7cm
\noindent August, 1996
\end{titlepage}

Recent developments in the type II superstring theory have shown that
$p$-brane solutions and strong-weak coupling dualities play an important
role in extracting the non-perturbative properties of the theory \cite
{HT,Wi1,DKL}. The supersymmetric $p$-brane solitonic solutions \cite{DKL,HS}
carrying R-R charges are described by open strings with Dirichlet boundary
conditions in the transverse directions and Neumann boundary conditions
in the $p+1$ world-volume directions \cite{DLP,Le,Gr,Po}. Through the
investigations of the $D$-$p$-brane by the conformal field theory approach
the soliton dynamics has been elucidated \cite{Wi2,Li,CK,Do,PCJ}. The action
 of the $D$-$p$-brane for $0\leq p\leq3$ has been constructed to have a
kinetic term of the Dirac-Born-Infeld (DBI) type \cite{Le} showing the
 coupling to  the NS-NS background fields and a Wess-Zumino term indicating
the coupling  to the R-R background fields as well as the NS-NS $2$-form
gauge field \cite{To,Scm,AS,Ts,GG,BR}.
The modified gauge-invariant field strength
provided by both the boundary gauge
field and the NS-NS $2$-form gauge field exists
in the two terms. The boundary gauge field is  the
open-string boundary condensate of the electromagnetic field.
 From the integration over the boundary gauge field for
the $D$-string and $D$-$2$-brane actions the fundamental string action
with tension given by Schwarz's formula \cite{Scw} and the bosonic part of
the eleven-dimensional supermembrane action \cite{BST} have been derived
respectively, by the Lagrangian formulation introducing the auxiliary
fields \cite{Scm} or by the canonical Hamiltonian approach \cite{AS}.
 The  $D$-string case provides support for the predictions
  of $SL(2,Z)$ duality of the type IIB theory.  From a
different point of view through the introduction of the Lagrange
multiplier field to make the semiclassical duality transformation for
the boundary gauge field these two actions have been reproduced and
further in the $D$-$3$-brane case the transformed action has turned out
to be the original form in terms of the dualized boundary gauge and
 background fields \cite{Ts}. It indicates that the $D$-$3$-brane action is
invariant under the $SL(2,R)$ duality transformations of the background
fields of type IIB theory combined with world-volume vector duality
in the four dimensions.
This duality invariance has been found in a similar way as the
four-dimensional non-linear electrodynamics \cite{GR}.

Here we will explore the $D$-$3$-brane in the canonical Hamiltonian
approach. Without resort to the Lagrange multiplier field we will try to
integrate the canonical boundary gauge field directly in the path
integral of the Hamiltonian form. The obtained effective action will get
 back to the original one in terms of dualized fields. We wil see how the
dual variable for the boundary gauge field appears in this direct
integration.

We start to write the $D$-$3$-brane action of type IIB theory
\begin{eqnarray}
S_D & = & \int d^{4}x[e^{-\phi}\sqrt{-\det(G_{\a\b}+\F_{\a\b})} \nonumber\\
 & & + \frac{1}{8}{\epsilon}^{\a\b\g\d}(\frac{1}{3}C_{\a\b\g\d}
 + 2C_{\a\b}{\cal F}_{\g\d} + C {\cal F}_{\a\b}{\cal F}_{\g\d})],
\label{S}\end{eqnarray}
which is the source term of the $D$-$3$-brane when combined with the type
IIB effective action. This closed-string effective action was presented
under some restriction on the R-R $4$-form gauge field
\cite{BHO,BBO} and was shown to be invariant under the $SL(2,R)$ duality
transformation of the type IIB theory \cite{HT,Wi1,Scw,BHO,SHW}. The kinetic
term is described by a dilaton $\phi$, a metric tensor $G_{\mu\nu}$ and an
antisymmetric tensor $B^{(1)}_{\mu\nu}$ in the NS sector. The embedding
 metric is given by $G_{\a\b}=\partial_{\a}X^{\mu}\partial_{\b}X^{\nu}
G_{\mu\nu} (\a, \b=0,\dots,3)$ where $X^{\mu}$ is the coordinates of the
ten-dimensional target space of type IIB theory. From the boundary gauge
field $A_{\a}$ to be mixed with the closed-string antisymmetric
$2$-form gauge field
the modified gauge-invariant field strength is defined as
 ${\cal F}_{\a\b}= F_{\a\b} - B_{\a\b}^{(1)}$
 with $F_{\a\b}=\partial_{\a}A
_{\b} - \partial_{\b}A_{\a}, B^{(1)}_{\a\b}=\partial_{\a}X^{\mu}
\partial_{\b}X^{\nu}B^{(1)}_{\mu\nu}$. The fields such as $C, C_{\a\b}\equiv
B^{(2)}_{\a\b}, C_{\a\b\g\d}$ are also the induced background R-R fields
 of type IIB theory. For simplicity we consider the case where the spacetime
 metric is Minkowskian. The DBI Lagrangian is expressed as
\beq
\sqrt{-\det(G_{\a\b}+\F_{\a\b})} =
  \sqrt{1 - {\E}^2 + {\B}^2 - ({\B}\cdot{\E})^2},
\label{DB}\eeq
where $E_i(i=1,2,3)$ stands for ${\cal F}_{0i}$ and $B_i$ for $\frac{1}{2}
{\epsilon}_{ijk}{\cal F}_ {jk}$. The canonical momenta conjugate to the
world-volume gauge fields $A_{\a}$ are given by
\beq
\pi_0=0, \mbox{\boldmath $\pi$} = -\frac{{\E}+({\B}\cdot{\E})
{\B}}{e^{\phi}\sqrt{D}} + {\C} + C{\B} ,
\label{pi}\eeq
where $C_i=\frac{1}{2}\epsilon_{ijk}C_{jk}$ and $\sqrt{D}$ denotes the
expression (\ref{DB}).  We pass to the Hamiltonian
\beq
H = p_iE_i - e^{-\phi}\sqrt{D} + (\partial_iA_0 + B^{(1)}_{0i})\pi_i
-(C_{0123} +C_{0i}B_i) ,
\label{Ha}\eeq
where
\beq
 p_i = \pi_i - C_i - CB_i .
\label{p}\eeq
In the non-linear equation (\ref{pi}) with (\ref{DB}), (\ref{p}) we will
derive a solution for {\E}. First by using (\ref{pi}) and (\ref{p}) we
get a relation
\beq
 {\E} = - \frac{e^{\phi}\sqrt{D}}{1 + {\B}^2} {\Q},
\label{E}\eeq
where
\beq
\left( \begin{array}{c} Q_1 \\ Q_2 \\ Q_3 \end{array} \right) = \left(
\begin{array}{ccc} 1+B_2^2+B_3^2 & -B_1B_2 & -B_1B_3 \\ -B_2B_1 & 1+B_3^2
+B_1^2 & -B_2B_3 \\ -B_3B_1 & -B_3B_2 & 1+B_1^2+B_2^2 \end{array} \right)
\left( \begin{array}{c} p_1 \\ p_2 \\ p_3 \end{array} \right).
\eeq
The next step is to substitute the relation into (\ref{DB}), which yields
consistently
\beq
 \sqrt{D} = \frac{(1 + {\B}^2)^{3/2}}{e^{\phi}\sqrt{M}}
\label{RD}\eeq
with $M=(1+{\B}^2)^2e^{-2\phi}+{\Q}^2+({\B}\cdot{\Q})^2$.
 Combining (\ref{E}) and (\ref{RD}) we obtain a solution ${\E}=-
\sqrt{(1+{\B}^2)/M}{\Q}$. Owing to the solution the combination $
p_iE_i-e^{-\phi}\sqrt{D}$ in (\ref{Ha}) turns out to be
$-\sqrt{(1+{\B}^2)}[e^{-2\phi}(1+{\B}^2)+{\Q}\cdot{\bp}]/
\sqrt{M}$.  As $M$ can be shown to be equal
to $(1+{\B}^2)[e^{-2\phi}(1+{\B}^2)
 +{\Q}\cdot{\bp}]$, we have
\beq
H = - \sqrt{e^{-2\phi}(1+{\B}^2)+{\Q}\cdot{\bp}}+(\partial_iA_0+
B^{(1)}_{0i})\pi_i -(C_{0123}+C_{0i}B_i).
\eeq

Now we are ready to write the partition function by using the Hamiltonian
form of the path integral as
\begin{eqnarray}
Z &=& \int \prod_{i=1}^{3}[d\pi_i]\prod_{\a=0}^{3}[dA_{\a}]
\exp[i\int d^4x(\pi_i\dot{A_i}-H(A,\pi))] \nonumber \\
 &=& \int [d(\pi,A)]\exp[i\int d^4x \{\partial_i\pi_iA_0+\pi_i
(\partial_0A_i-B^{(1)}_{0i})+\sqrt{N}+C_{0123}+C_{0i}B_i\} ],
\label{PI}\end{eqnarray}
where the integration over $\pi_0$ is canceled by the gauge group volume and
$N=(1+{\B}^2)(e^{-2\phi}+{\bp}^2)-({\bp}\cdot{\B})^2$. The
integration over $A_0$ brings a $\d$-function $\d(\partial_i\pi_i)$ down
to the path integral, which allows us to write $\pi_i=\epsilon_{ijk}
\partial_jD_k$ with a vector field $D_i$. In the integration over $A_i$
the magnetic fields $B^{0}_i$ defined by $B^{0}_i=
{\epsilon}_{ijk}{\partial}_jA_k$ are not independent fields because of
${\partial}_iB^{0}_i=0$. We make change of integration variables from
 $A_i$ to $B^{0}_i$ and treat the magnetic fields $B^{0}_i$
as the independent variables but with
a constraint ${\partial}_iB^{0}_i=0$, whose $\d$-function is expressed
by introducing an auxiliary field $D_0$ as
\beq
 \prod_x\d(\partial_i
B^{0}_i(x))=\int [dD_0]\exp(-i\int d^4xD_0\partial_iB^{0}_i).
\eeq
Then the partition function is described by
\beq
Z=\int\prod_{\a=0}^{3}[dD_{\a}]\prod_{i=1}^{3}[dB^{0}_i]\exp[i\int d^4x
\{\sqrt{N}-B_i(D_{0i}-C_{0i}) + \frac{1}{8}\epsilon^{\a\b\g\d}(
\frac{1}{3}C_{\a\b\g\d}-2D_{\a\b}B^{(1)}_{\g\d}) \} ]
\label{Z}\eeq
with $D_{\a\b}=\partial_{\a}D_{\b}-\partial_{\b}D_{\a}$. Through a shift
the integration over $B^{0}_i$ becomes that over $B_i$.

Here we make the integration over $B_i$ by using the saddle point
approximation. It is noted that
\beq
N = (1+{\B}^2)e^{-2\phi} + ({\cp} - C{\B})^2 +
({\cp}\times{\B})^2,
\label{N1}\eeq
where $\p_i=\frac{1}{2}\epsilon_{ijk}(D_{jk}-C_{jk})$.
The variation with respect to $B_i$ yields the saddle point equation
\beq
\left(\begin{array}{ccc} \Lambda + \p_2^2 + \p_3^2 & -\p_1\p_2 & -\p_1\p_3
\\ -\p_2\p_1 & \Lambda + \p_3^2 + \p_1^2 & -\p_2\p_3 \\
   -\p_3\p_1 & -\p_3\p_2 & \Lambda + \p_1^2 + \p_2^2 \end{array} \right)
\left( \begin{array}{c} B_1 \\ B_2 \\ B_3 \end{array} \right) =
\left( \begin{array}{c} \sqrt{N}d_{01} + C\p_1 \\ \sqrt{N}d_{02} + C\p_2 \\
\sqrt{N}d_{03} + C\p_3 \end{array} \right),
\eeq
where $\Lambda=e^{-2\phi} + C^2, d_{0i}=D_{0i} - C_{0i}$. We use the
inversion of the matrix
\beq
\frac{1}{\Lambda(\Lambda+{\cp}^2)}\left( \begin{array}{ccc}
\Lambda + \p_1^2 & \p_1\p_2 & \p_1\p_3 \\ \p_2\p_1 & \Lambda + \p_2^2 &
\p_2\p_3 \\ \p_3\p_1 & \p_3\p_2 & \Lambda + \p_3^2 \end{array} \right)
\eeq
to have
\beq
{\B}= \frac{1}{\Lambda(\Lambda + {\cp}^2)}[\sqrt{N} \{ \Lambda
{\bd} + ({\cp}\cdot{\bd}){\cp} \} + (\Lambda + {\cp}^2)C\cp ]
\label{B}\eeq
with $d_i=d_{0i}$. Substituting this non-linear relation back into
(\ref{N1}) to observe the cancellation of the $\sqrt{N}$ term,
we can determine $N$ consistently as
\beq
N=\frac{e^{-2\phi}(\Lambda + {\cp}^2)^3}{\Lambda(\Lambda +
{\cp}^2)^2 - \Lambda^2{\bd}^2 - ({\cp}\cdot{\bd})^2
(2\Lambda + {\cp}^2) - \Lambda({\cp}\times{\bd})^2}.
\label{N2}\eeq
In this self-consistent way we find the explicit solution (\ref{B})
accompanied with (\ref{N2}) for the non-linear saddle point equation.
Plunging the complicated solution into the action in (\ref{Z}), we have
\begin{eqnarray}
\int d^4x &[ &  \frac{\Lambda(\Lambda + {\cp}^2) - \Lambda {\bd}^2
- ({\cp}\cdot{\bd})^2 }{\Lambda(\Lambda + {\cp}^2 )} \sqrt{N}
\nonumber \\  & + & \frac{1}{8} \epsilon^{\a\b\g\d}
 (\frac{1}{3} C_{\a\b\g\d}
-2B^{(1)}_{\a\b}D_{\g\d} - \frac{C}{\Lambda} \D_{\a\b}\D_{\g\d}) ],
\end{eqnarray}
where ${\cal D}_{\a\b}=D_{\a\b}-C_{\a\b}$. Further $N$ can be so rewritten as
\beq
N = \frac{e^{-2\phi}(\Lambda + {\cp}^2)^2}{ \Lambda(\Lambda + {\cp}^2)
 - \Lambda{\bd}^2 - ({\cp}\cdot{\bd})^2}
\label{N3}\eeq
that for the kinetic part we derive a compact expression $e^{-\phi}
\sqrt{-\det(\sqrt{\Lambda}\eta_{\a\b} + \D_{\a\b})}/\Lambda$, which is just
of the DBI type. This result producing the Lorentz invariant expression
 can be generalized to the general metric $G_{\a\b}$ as
\begin{eqnarray}
Z&=&\int \prod_{\a=0}^{3}[dD_{\a}]\exp[i\int d^4x \{ e^{-\tilde{\phi}}
\sqrt{-\det(\tilde{G}_{\a\b}+{\cal D}_{\a\b})} \nonumber
\\ & +& \frac{1}{8}\epsilon^{\a\b\g\d}(\frac{1}{3}C_{\a\b\g\d}
+ 2 \tilde{C}_{\a\b}{\cal D}_{\g\d}
 + \tilde{C}{\cal D}_{\a\b}{\cal D}_{\g\d} ) \} ],
\end{eqnarray}
where ${\cal D}_{\a\b}= D_{\a\b}- \tilde{B}^{(1)}_{\a\b}$ and
\begin{eqnarray}
\tilde{B}^{(1)}_{\a\b}=C_{\a\b}
,& \tilde{C}_{\a\b}=-B^{(1)}_{\a\b} , \nonumber \\
e^{-\tilde{\phi}}=\frac{e^{-\phi}}{\Lambda}=\frac{1}{e^{-\phi}
+ e^{\phi}C^2} , & \tilde{C}= - \displaystyle\frac{C}{\Lambda} = -
\displaystyle\frac{Ce^{\phi}}{e^{-\phi} + e^{\phi}C^{2}},  \\
\tilde{G}_{\a\b} = \sqrt{\Lambda}G_{\a\b} = e^{(\tilde{\phi}-\phi)/2}
G_{\a\b}.  \nonumber
\end{eqnarray}
The obtained relations in $(21)$ show the basic inversion $\lambda
 \rightarrow - 1/{\lambda}$ in the $SL(2,R)$ duality transformation
, which is specified by $p=0, q=1, r=-1, s=0$ for
\begin{eqnarray}
g_{\a\b} \rightarrow g_{\a\b},& \lambda \rightarrow (p\lambda + q)/
(r\lambda + s) , \nonumber \\ B^{(1)}_{\a\b} \rightarrow s B^{(1)}
 - r C_{\a\b}, & C_{\a\b} \rightarrow pC_{\a\b} - qB^{(1)}_{\a\b}
\end{eqnarray}
with $g_{\a\b} =e^{-\phi/2}G_{\a\b}, \lambda= C + ie^{-\phi}$.
Although $D_i$ and $D_0$ are separately induced, the final result shows
that they are combined into a four-dimensional dual vector. This dualized
 action was presented in the Lagrangian formulation for the duality
transformation  by adding the
Lagrange multiplier term $\frac{1}{2}\Lambda_{\a\b}({\cal F}_{\a\b}
-2\partial_{\a}A_{\b} +B^{(1)}_{\a\b})$ to the $D$-$3$-brane action
to symplify the integration over the boundary gauge field and further
integrating over $\F_{\a\b}$ in a special Lorentz frame where $\F_{\a\b}$
is block-diagonal with eigenvalues $f_1,f_2$ by means of the saddle
point approximation for $f_1,f_2$ \cite{Ts}. In our Hamiltonian approach
without choosing such a special Lorentz frame we have performed the
${\cal F}_{ij}$ integration in the arbitrary Lorentz frame to derive the
dualized action. Together with the simple invariance under the shift of
$C$ we have given the alternative way to show the $D$-$3$-brane action is
invariant under the full $SL(2,R)$ duality transformation.

Here we return to the path integral (\ref{PI}) in the phase space.
Alternatively if the integration with respect to the canonical momentum
$\pi_i$ is evaluated also by the saddle point value, which is in the
similar way given by
\beq
\mbox{\boldmath $\pi$} =-\frac{\sqrt{N'}}{1+\B^2} ( \E + (\B \cdot \E)\B)
+ \C +C\B
\label{pi2}\eeq
with
\beq
N' = \frac{e^{-2\phi}(1+\B^2)^2}{1-\E^2 +\B^2 -(\B\cdot \E)^2},
\label{NP}\eeq
then we go back to the starting classical $D$-$3$-brane action itself
$(\ref{S})$ as expected. It is observed that these expressions (\ref{pi2}),
(\ref{NP}) show similar behaviors to (\ref{B}), (\ref{N3}), respectively.

 In the Hamiltonian approach to the $D$-$3$-brane
we have seen that the constraint $\partial_i\pi_i=0$ provided from the
$A_0$ integration brings about the dual variable $D_i$ as $\pi_i=
\epsilon_{ijk}\partial_jD_k$, while for the $D$-string and $D$-$2$-brane
cases  each canonical momentum is described
by $\pi=q_1$ and $\pi_i=\epsilon_{ij}\partial_jy$ where $q_1$ is a
charge of the dyonic string solution in type IIB theory and the scalar
function $y$ becomes an additional target space coordinate in the
eleven-dimensional target space of the fundamental $2$-brane \cite{AS}.
In transforming the variables from $A_i$ to $F_{ij}$ an extra dual variable
$D_0$ has appeared, which is a conspicuous behavior in the $D$-$3$-brane
case in contrast to the $D$-string and $D$-$2$-brane cases.
Moreover when integrating with respect to the magnetic field $F_{ij}$
we have met the non-linear saddle point equation, which is
compared with the linear ones in the
$D$-string and $D$-$2$-brane cases. In our Hamiltonian approach
the canonical momentum $\pi_{\a}$ has been replaced by the dual variable
$D_{\a}$ and this change of variable has been generated as a result of the
integration over $A_{\a}$.

In conclusion, starting from the $D$-$3$-brane action we have constructed the
Hamiltonian by solving the non-linear equation in order to obtain the
electric field of the boundary gauge field and integrated over the
magnetic field by solving the non-linear saddle point equation in the
path integral of the Hamiltonian form to derive the effective action,
 that is just the dualized action in terms of the dual gauge field
and the basic $SL(2,R)$ transformed background fields of the type IIB
theory. The essential ingredient in this Hamiltonian approach is to
look for the explicit solutions of these two non-linear equations.
They have been solved analogously in a self-consistent manner.
Although the derived solutions are highly involved, the resulting
effective action is put in the compact form. In the transformation of
the integration variables from the three space-components of the boundary
gauge field to the magnetic fields we need a new field, which turns into
a time-component of the dual gauge field and plays an important role to
obtain the Lorentz-invariant expression of the dualized action.
Our demonstration of the self-duality of the $D$-$3$-brane action is a
direct approach in the sense that the extraneous structure as the Lagrange
multiplier field does not need to be introduced from the starting point.
We hope that our prescription sheds insight into the structure of duality
transformation from the Hamiltonian point of view in the phase space
of the theory.

\end{document}